\documentclass{aa}
\usepackage[varg]{txfonts}

\usepackage{natbib}
\bibpunct{(}{)}{;}{a}{}{,} 

\usepackage{graphicx}

\begin{document}

\title{Is 4C$+$29.48 a $\gamma$-ray source?} 

\author{K. \'E. Gab\'anyi\inst{1,2}, S. Frey\inst{1}, T. An\inst{3}}

\offprints{K. \'E. Gab\'anyi, \email{gabanyi@konkoly.hu}}

\institute{Konkoly Observatory, MTA Research Centre for Astronomy and Earth Sciences, Konkoly Thege Mikl\'os \'ut 15-17, H-1121 Budapest, Hungary \and MTA-ELTE Extragalactic Astrophysics Research Group, ELTE TTK P\'azm\'any P\'eter s\'et\'any 1/A, H-1117, Budapest, Hungary \and Shanghai Astronomical Observatory, Key Laboratory of Radio Astronomy, Chinese Academy of Sciences, 80 Nandan Road, 200030 Shanghai, P. R. China}

\date{Received / Accepted}

\abstract{The {\it Fermi} Large Area Telescope revealed that the extragalactic $\gamma$-ray sky is dominated by blazars, active galactic nuclei (AGN) whose jet is seen at very small angle to the line of sight. To associate and then classify the $\gamma$-ray sources, data have been collected from lower frequency surveys and observations. Since those have superior angular resolution and positional accuracy compared to the $\gamma$-ray observations, some associations are not straightforward.}
{The $\gamma$-ray source 3FGL\,J1323.0+2942 is associated with the radio source 4C$+$29.48 and classified as a blazar of unknown type, lacking optical spectrum and redshift. The higher-resolution radio data showed that 4C$+$29.48 comprises three bright radio-emitting features located within a $\sim 1'$-diameter area. We aim to reveal their nature and pinpoint the origin of the $\gamma$-ray emission.}
{We (re-)analyzed archival Very Large Array (VLA) and unpublished very long baseline interferometry (VLBI) observations conducted by the Very Long Baseline Array (VLBA) and the European VLBI Network of 4C$+$29.48. We also collected data form optical, infrared and X-ray surveys.}
{According to the VLBI data, the northernmost complex of 4C$+$29.48 contains a blazar with a high brightness temperature compact core and a steep-spectrum jet feature. The blazar is positionally coincident with an optical source at a redshift of $1.142$. Its mid-infrared colors also support its association with a $\gamma$-ray emitting blazar. The two other radio complexes have steep radio spectra similar to AGN-related lobes and do not have optical or infrared counterparts in currently available surveys. Based on the radio morphology, they are unlikely to be related to the blazar. There is an optical source between the two radio features, also detected in infrared wavebands. We discuss the possibilities whether the two radio features are lobes of a radio galaxy, or gravitationally lensed images of a background source.}
{We propose to associate the $\gamma$-ray source 3FGL\,J1323.0+2942 in subsequent versions of the {\it Fermi} catalog with the blazar residing in northernmost complex. We suggest naming this radio source J1323$+$2941A to avoid misinterpretation arising from the fact that the coordinates of the currently listed radio counterpart 4C$+$29.48 is closer to a most probably unrelated radio source.}


\maketitle

\section{Introduction}

Most of the active galactic nuclei (AGN) detected by the {\it Fermi} Large Area Telescope (LAT) survey in $\gamma$-rays are classified as blazars \citep{Fermi3_AGNcat}, radio-loud AGN whose jet points at very small angle to the line of sight \citep{UrryPad}. The $\gamma$-ray emitting non-blazar type AGN ($2$ \%) include a few radio galaxies, compact symmetric objects, often regarded as young radio galaxies \citep{Fanti_CSO}, and narrow-line Seyfert 1 sources. The classification of the different types of $\gamma$-ray AGN is done by 
looking for the low-energy counterpart associated with the respective $\gamma$-ray source.
The positional uncertainty of {\it Fermi} LAT is much larger than that of the lower-energy surveys used to find counterpart candidates, thus associations are not always straightforward and a few duplicate associations exist \citep{Massaro_Fermi_review}. 

Approximately one-third of the blazars in the catalog of \cite{Fermi3_AGNcat} are of uncertain type. Their counterparts either lack optical spectrum, or the quality of their optical spectra does not allow to distinguish between the two flavors of blazars, flat-spectrum radio quasar (FSRQ) and BL Lac object, which are traditionally distinguished by the equivalent width of their emission lines \citep[e.g.,][]{BLLac_FSRQ}.

The spectral energy distribution (SED) of blazars is dominated by the non-thermal emission of their jets and are characterized by two bumps. The lower-energy bump of the SED (peaking between millimeter and X-ray regimes) originates from the synchrotron emission of the electrons and/or positrons in the jet. The higher-energy bump in the $\gamma$-ray regime is usually attributed to inverse Compton emission in the framework of leptonic models \citep[e.g., ][and references therein]{Giommi2012} or to synchrotron emission from heavy charged particles, for example, protons in hadronic models. In the leptonic models, the seed photons originate either from the synchrotron jet \citep[synchrotron self-Compton scenario, e.g.,][]{SSC}, or from external sources, such as the accretion disk, the broad-line region \citep{BLR_gamma}, and/or the torus \citep[external Compton scenario, e.g.,][]{externalCompton}. Synchrotron self-Compton process is usually invoked to describe the spectral energy distribution of BL Lac objects, the external Compton mechanism is often assumed in the case of FSRQs \citep{Massaro_Fermi_review}.

The $\gamma$-ray source 3FGL J1323.0$+$2942 is associated with the radio source 4C$+$29.48, and listed as a blazar candidate of uncertain type in the latest, third {\it Fermi} LAT catalog \citep[3FGL, ][]{FermiLAT3}. According to the third catalog of {\it Fermi} LAT detected AGN \citep{Fermi3_AGNcat}, its counterpart lacks an optical spectrum. 

The source was not detected by the VERITAS imaging atmospheric Cherenkov telescope between 2007 and 2012 in the TeV regime \citep{VERITAS_non}. The source is listed as an FSRQ based upon the paper of \cite{Cornwell1986} (see below). \cite{VERITAS_non} calculated the VERITAS flux upper limit and found it larger than the extrapolated {\it Fermi} LAT flux measurements. Because of the unknown redshift of the source, they assumed redshifts of $z=0.1$ and $z=0.5$, when taking the electron-positron pair production with photons of the extragalactic background light into account.

The source 4C$+$29.48 has an elongated structure in the NRAO Very Large Array Sky Survey \citep[NVSS,][]{nvss} image, which was used to cite the radio flux density of the source in the catalog of \cite{Fermi3_AGNcat}. However, it is resolved into three radio sources positioned roughly along a line according to the finer-resolution Faint Images of the Radio Sky at Twenty-Centimeters survey \citep[FIRST,][]{first}. This triplet of radio-emitting features was already studied in the 1980s by \cite{Cornwell1986} using radio observations with the Very Large Array (VLA). Following their terminology, we will refer to the three sources from north to south as Complex A, B, and C (see Fig. \ref{fig:FIRST_Fermi}).

\cite{VLBA_incorrect_pointing_Lico} conducted Very Long Baseline Array (VLBA) observations at $5$\,GHz of a sample of $\gamma$-ray sources including 4C$+$29.48. They reported that the source was not detected and it is excluded from their analysis due to incorrect pointing. \cite{VLBI_imaging_Morgan} performed very long baseline interferometry (VLBI) observation at $8$\,GHz of the field of 4C$+$29.48 using an ad-hoc array consisting of four European stations. They selected the field of 4C$+$29.48 to illustrate the capability of the DiFX software correlator \citep{deller} in wide-field imaging. They checked the positions of the three complexes and also mapped the entire primary beam of the array, within a radius of $90''$ around Complex B. They were only able to detect the northernmost feature, Complex A. 

To understand the nature and pinpoint the location of the $\gamma$-ray emitting source we collected and analyzed archival radio data of 4C$+$29.48. We introduce these data sets and describe the data reduction in Sect. \ref{sec:data}. We present the results of the observations in Sect. \ref{sec:res}. In Sect. \ref{sec:disc}, we discuss our findings in the light of multiwavelength information. Finally, in Sect. \ref{sec:sum}, we give a summary.
We assumed a flat $\Lambda$CDM cosmological model with $H_0=70\mathrm{\,km\,s}^{-1}\mathrm{\,Mpc}^{-1}$, $\Omega_\mathrm{m}=0.27$, and  $\Omega_\Lambda=0.73$.

\section{Archival radio data} \label{sec:data}

\subsection{Very Large Array data}

\cite{Saikia1984} conducted VLA C-configuration observations of 4C$+$29.48 on 1980 July 13 at $4.89$, and $15$\,GHz (project code: SINH). The on-source integration times were $\sim27$\,min and $\sim29$\,min, respectively. \cite{Cornwell1986} conducted VLA B-configuration observations of 4C$+$29.48 on 1982 September 12 at $1.47$, and at $4.89$\,GHz (project code: PP), and VLA A-configuration observations on 1983 November 25 at three frequencies, at $1.49$, $4.86$, and $14.94$\,GHz (project code: AS135). In each of the C and B configuration observations one intermediate-frequency channel (IF) was used with $50$\,MHz bandwidth, while in the A configuration observations, two IFs were used each with $50$\,MHz bandwidth. In the B configuration observations, the on-source integration times were $\sim 34$\,min at $1.47$\,GHz and $\sim 30$\,min at $4.89$\,GHz. In the A configuration observations, the on-soure integration times were $\sim 12$\,min at $1.49$\,GHz and at $14.94$\,GHz, and $8$\,min at $4.86$\,GHz. 

We downloaded the data from the US National Radio Astronomy Observatory (NRAO) archive\footnote{The 1980 observations were among the first conducted with the completed VLA. They could be found in the NRAO archive only by searching for the date of the observations and not by source name or coordinates.} and reduced it using the NRAO Astronomical Image Processing System \cite[{\sc aips},][]{aips} in a standard way. 3C\,286 was used as the primary flux density calibrator. Standard {\sc aips} tasks were used to calibrate the data sets\footnote{http://www.aips.nrao.edu/cook.html}. For imaging, we used the {\sc difmap} software \citep{difmap}.

\subsection{Very Long Baseline Interferometry observations}

\begin{table*}
\caption{Details of archival VLBI observations targeting 4C$+$29.48. In Col. 5, we list the pointing coordinates.}
\label{tab:VLBI_obs}
\centering
\begin{tabular}{cccccc}
\hline \hline
Date & Array & Frequency & Project code & Pointing coordinates & Reference \\
 & & (GHz) & & Right ascension, Declination & \\
\hline
2007.12.12 & EfMaMcWz & 8.4\tablefootmark{a} & - & $13^\textrm{h} 23^\textrm{m}02\fs75$, $+29\degr 41\arcmin 32\farcs5$ & \cite{VLBI_imaging_Morgan} \\
2013.04.16 & EVN & 1.6\tablefootmark{a} & EL034A &  $13^\textrm{h} 23^\textrm{m}02\fs587$, $+29\degr 41\arcmin 33\farcs360$ & EVN archive\\
2013.04.28 & VLBA & 4.3, 7.6 & BP171 & $13^\textrm{h} 23^\textrm{m}00\fs873$, $+29\degr 41\arcmin 44\farcs812$ & astrogeo.org \\
2013.05.07 & VLBA & 7.6\tablefootmark{b}  & S4195A &  $13^\textrm{h} 23^\textrm{m}00\fs873$, $+29\degr 41\arcmin 44\farcs812$ & astrogeo.org\\
2013.05.18 & VLBA & 7.6 & S4195B & $13^\textrm{h} 23^\textrm{m}00\fs873$, $+29\degr 41\arcmin 44\farcs812$ & astrogeo.org \\
2013.06.22 & VLBA & 7.6 & S4195C & $13^\textrm{h} 23^\textrm{m}00\fs873$, $+29\degr 41\arcmin 44\farcs812$ & astrogeo.org \\
2013.09.30 & VLBA & 5\tablefootmark{a} & S6340D & $13^\textrm{h} 23^\textrm{m}02\fs59$, $+29\degr 41\arcmin 33\farcs4$\tablefootmark{c} & \cite{VLBA_incorrect_pointing_Lico} \\ 
2014.01.24 & VLBA & 4.87 & BL189 &  $13^\textrm{h} 23^\textrm{m}02\fs357$, $+29\degr 41\arcmin 34\farcs270$ & NRAO archive \\
\end{tabular}
\tablefoot{\tablefoottext{a}{Phase-referenced observations.}\tablefoottext{b}{Dataset could not be imaged.}\tablefoottext{c}{The listed coordinates are from \cite{VLBA_incorrect_pointing_Lico} and not directly derived from the data as in the other cases.}}
\end{table*}

Searching the VLBI image database of the astrogeo.org website\footnote{http://astrogeo.org/ maintained by L. Petrov}, we found four $7.6$\,GHz and one $4.3$\,GHz VLBA observations of the radio source J1323$+$2941. The coordinates of the source are right ascension $13^\textrm{h}23^\textrm{m}00\fs8735$ and declination $+29\degr 41\arcmin 44\farcs812$ with an accuracy of $0.25$\,mas, thus it is coincident with the position of Complex A seen in the FIRST image (Fig. \ref{fig:FIRST_Fermi}). The source was observed within the framework of the VLBA Calibrator Survey--7 project (BP171) and the VLBA follow-up of {\it Fermi} sources (S4195; Y. Y. Kovalev et al. 2018, in prep.).

We downloaded the calibrated visibility files to perform imaging and self-calibration using the {\sc difmap} software. In all observations the observing setup was similar, eight separate IFs each with a bandwidth of $32$\,MHz were used. All the ten 25-m antennas of the VLBA participated in the observations on 2013 April 28. On 2013 May 7, May 18, and June 22 one (Pie Town), two (Fort Davis and Saint Croix), and one antenna (Fort Davis) were missing, respectively. On 2013 April 28, the on-source times at both frequencies were $40$\,s. On 2013 May 18 and 2013 June 22, the on-source times were $\sim 44$\,min and $\sim 6$\,min, respectively. We were not able to image the $7.6$-GHz dataset taken on 2013 May 7, but the on-source time was very short, only $\sim 10$\,s. We were able to image however the three other 7.6-GHz datasets which were observed within two months, between 2013 April 28 and 2013 June 22. Since there was no significant difference between the obtained images and flux densities, we combined the three observations to reduce the noise level when creating the final image of the source.

The source 4C$+$29.48 was also observed with the European VLBI Network (EVN) on 2013 April 15 at 1.6\,GHz (project code: EL043A, PI: R. Lico)  in phase-reference mode \citep[e.g.,][]{phase-ref}. The observation targeted Complex B.
The following antennas provided data: Effelsberg (Germany), Medicina (Italy), Onsala (Sweden), Toru\'n (Poland), and the Westerbork Synthesis Radio Telescope (WSRT, the Netherlands). The observation was conducted in e-VLBI mode \citep{eEVN}. In that mode, the antennas are connected to the central EVN data processor at the Joint Institute for VLBI in Europe (JIVE, Dwingeloo, the Netherlands) via optical fibre networks to allow for real-time correlation. The total bandwidth was $128$\,MHz in both left and right circular polarizations, and the maximum data transmission rate was $1024\mathrm{\,Mbit\,s}^{-1}$. The on-source integration time was $\sim 18$\,min.
We downloaded the dataset and reduced it following the same steps as for experiment EG070A (an observation conducted in the same e-EVN session) reported in \cite{ngc5515}. 

According to the NRAO archive, 4C$+$29.48 was included in the project BL189 (PI: J. Linford) and observed on 2014 January 24 at $4.87$\,GHz. The array consisted of the ten antennas of the VLBA. Eight separate IFs each with $32$\,MHz bandwidth were used. The on-source time was $296$\,min. This observation was not phase-referenced. The pointing coordinates given for 4C$+$29.48 were right ascension $13^\textrm{h} 23^\textrm{m}02\fs3570$ and declination $+29\degr 41\arcmin 34\farcs270$, thus it was close to Complex B 
similarly to the EVN observation. The calibrator sources were J1310+3220, OQ208, and 3C286. To reduce the data we used standard {\sc aips} tasks. During fringe-fitting, fringes were found for 4C$+$29.48 with delays of a few hundred nanoseconds reaching even microseconds in a few cases. In contrast, for the calibrators the delays ranged from several hundred picoseconds to few nanoseconds. The distance ($\sim 21''$) and direction (northwestern) implied by the baseline-dependent delays confirm that the found fringes originate from Complex A.

The details of the archival VLBI observations are summarized in Table \ref{tab:VLBI_obs}. For comparison we list the already published observations of \cite{VLBI_imaging_Morgan} and \cite{VLBA_incorrect_pointing_Lico}. The array used by \cite{VLBI_imaging_Morgan} consisted of the following telescopes: Effelsberg (Germany), Matera (Italy), Medicina (Italy), and Wettzell (Germany).

\section{Results} \label{sec:res}

\subsection{VLA observations}

\begin{figure}
\centering
\resizebox{\hsize}{!}{\includegraphics[clip]{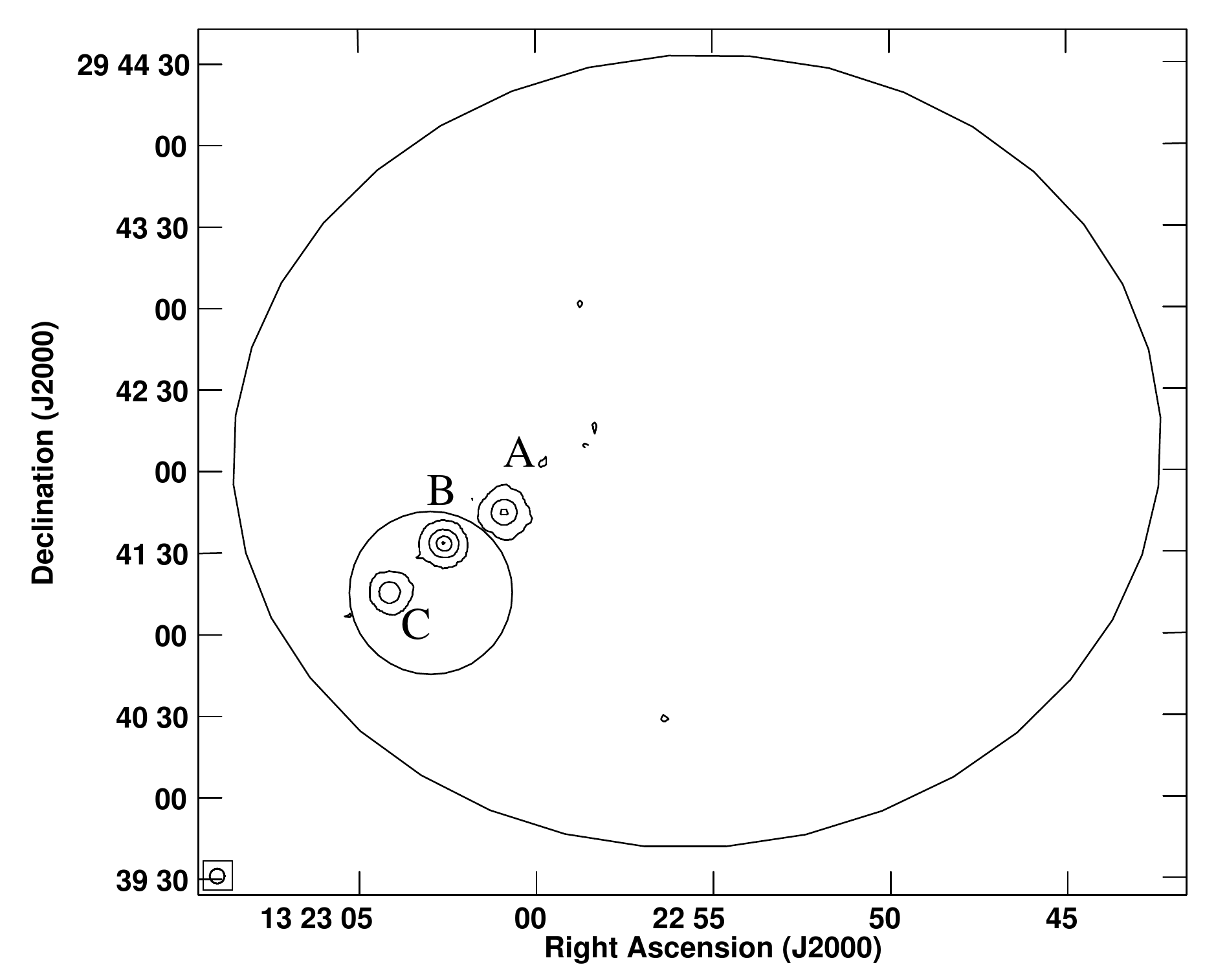}}
\caption{$1.4$-GHz FIRST image of 4C$+$29.48. Peak intensity is $614.7\textrm{\,mJy\,beam}^{-1}$, lowest contour levels at $8 \sigma$ image noise level, $1.3 \textrm{\,mJy\,beam}^{-1}$. Further contour levels are $(50, 250, 450) \times 1.3 \textrm{\,mJy\,beam}^{-1}$. Beam size is $5\farcs4 \times 5\farcs4$. The ellipse represents the position of the {\it Fermi} $\gamma$-ray source \citep{FermiLAT3}, the circle represents the position of the {\it ROSAT} X-ray source \citep{rosat_Boller} at $95$\,per cent confidence level.}
\label{fig:FIRST_Fermi}
\end{figure}

\begin{figure}
\centering
\resizebox{\hsize}{!}{\includegraphics[bb=0 0 530 365, clip]{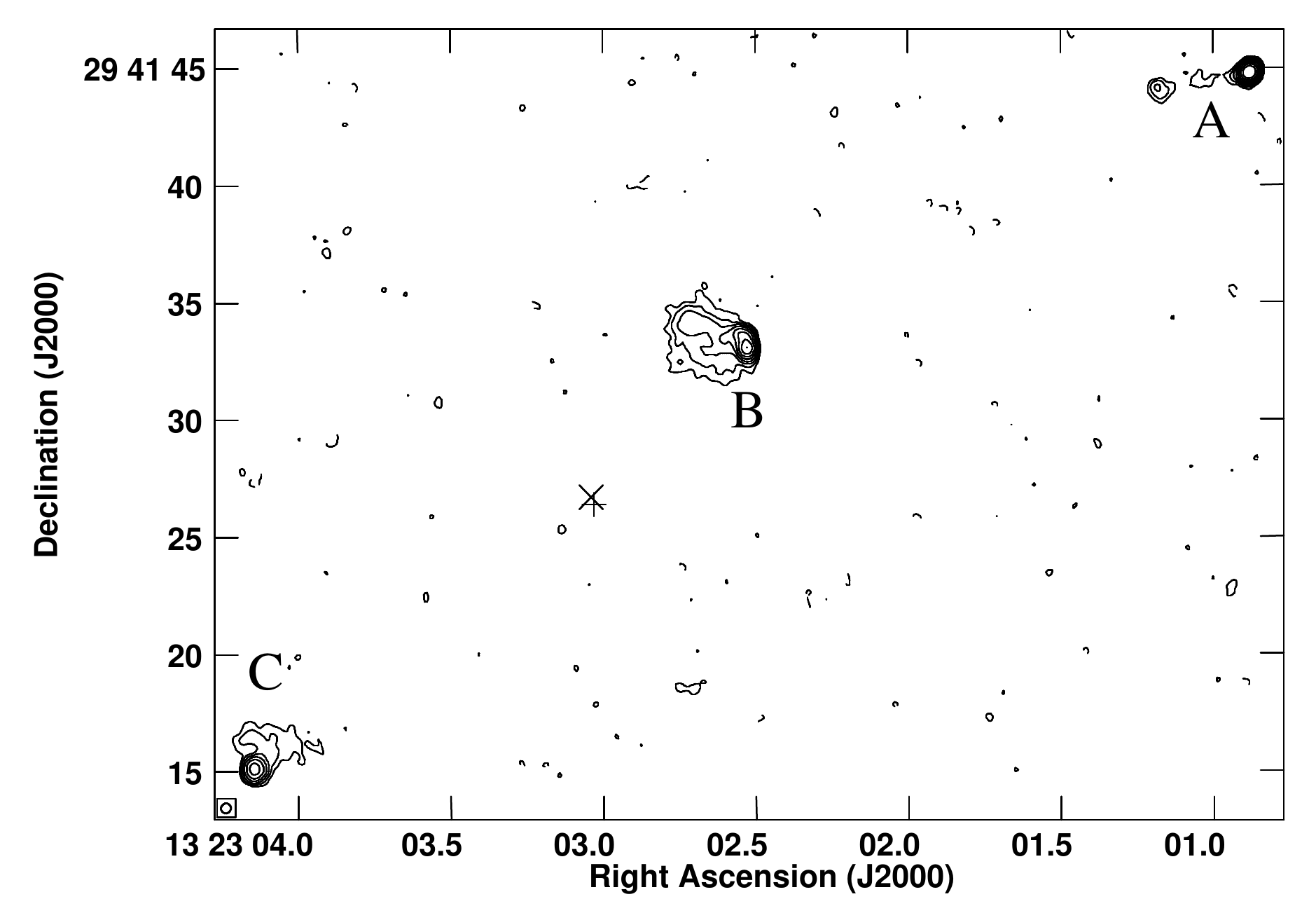}} 
\caption{$4.86$-GHz VLA-A map. Peak intensity is $211 \mathrm{\,mJy\,beam}^{-1}$, lowest contour levels at $3 \sigma$ rms noise level, $0.6 \mathrm{\,mJy\,beam}^{-1}$. Further contour levels increase by a factor of two. Beam size is $0\farcs43 \times 0\farcs41$ at a position angle of $-39\degr$. The plus sign indicates the position of the faint optical source, SDSS\,J132303.03$+$294126.4. The cross indicates the position of the infrared source WISE J132303.0$4+$294126.7 associated with that optical source. The positional accuracies of these measurements are much smaller than their representative symbols.}
\label{fig:VLAA_Cband}
\end{figure}

\begin{figure}
\centering
\resizebox{\hsize}{!}{\includegraphics[bb=13 8 530 330, clip]{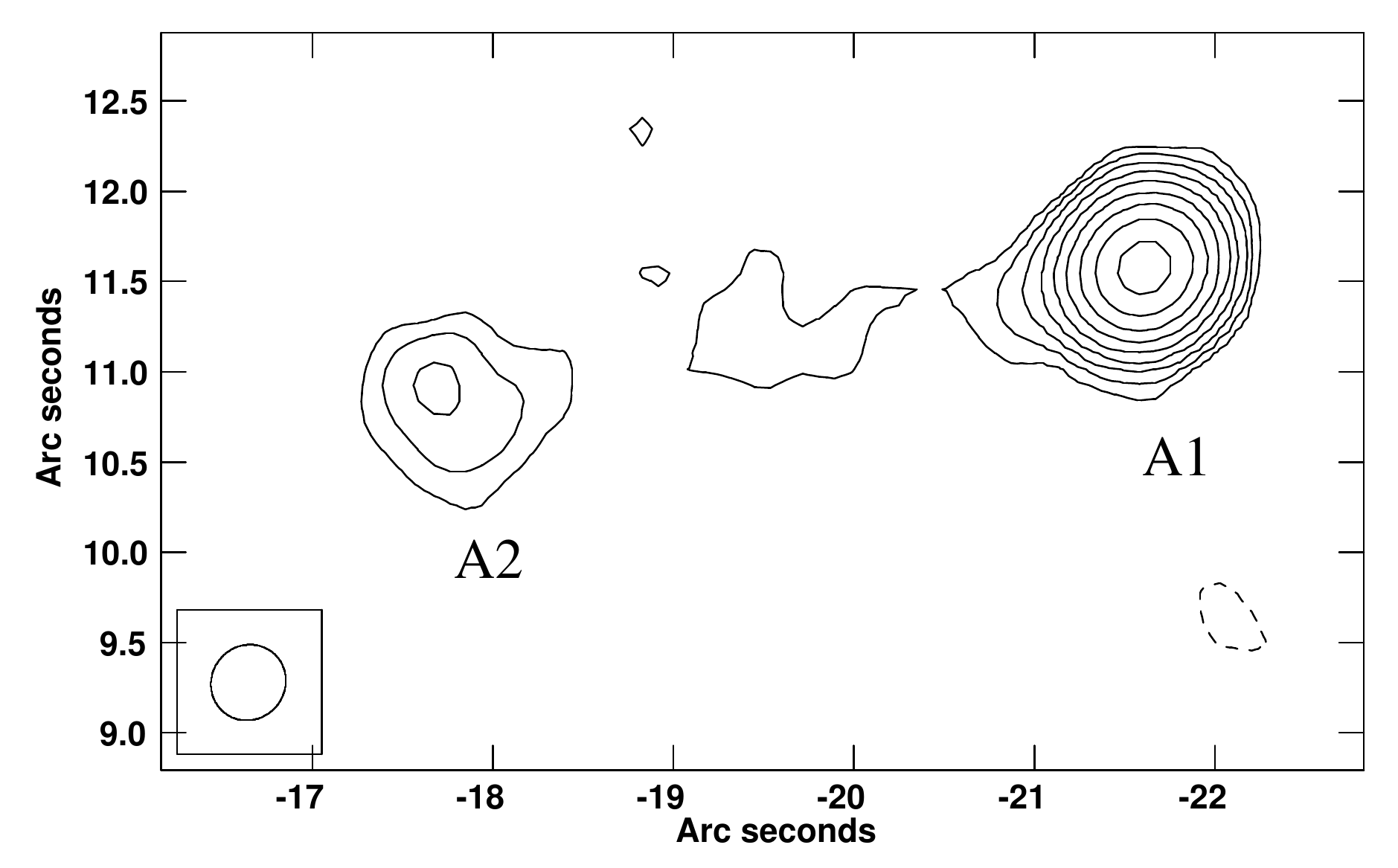}}
\caption{$4.86$-GHz VLA-A map of Complex A. Peak intensity is $211 \mathrm{\,mJy\,beam}^{-1}$, lowest contour levels at $3 \sigma$ rms noise level, $0.6 \mathrm{\,mJy\,beam}^{-1}$. Further contour levels increase by a factor of two. Beam size is $0\farcs43 \times 0\farcs41$ at a position angle of $-39\degr$ shown at the lower left part of the image.}
\label{fig:A_VLAA_Cband}
\end{figure}

\begin{figure}
\centering
\resizebox{\hsize}{!}{\includegraphics[bb=50 220 565 570, clip]{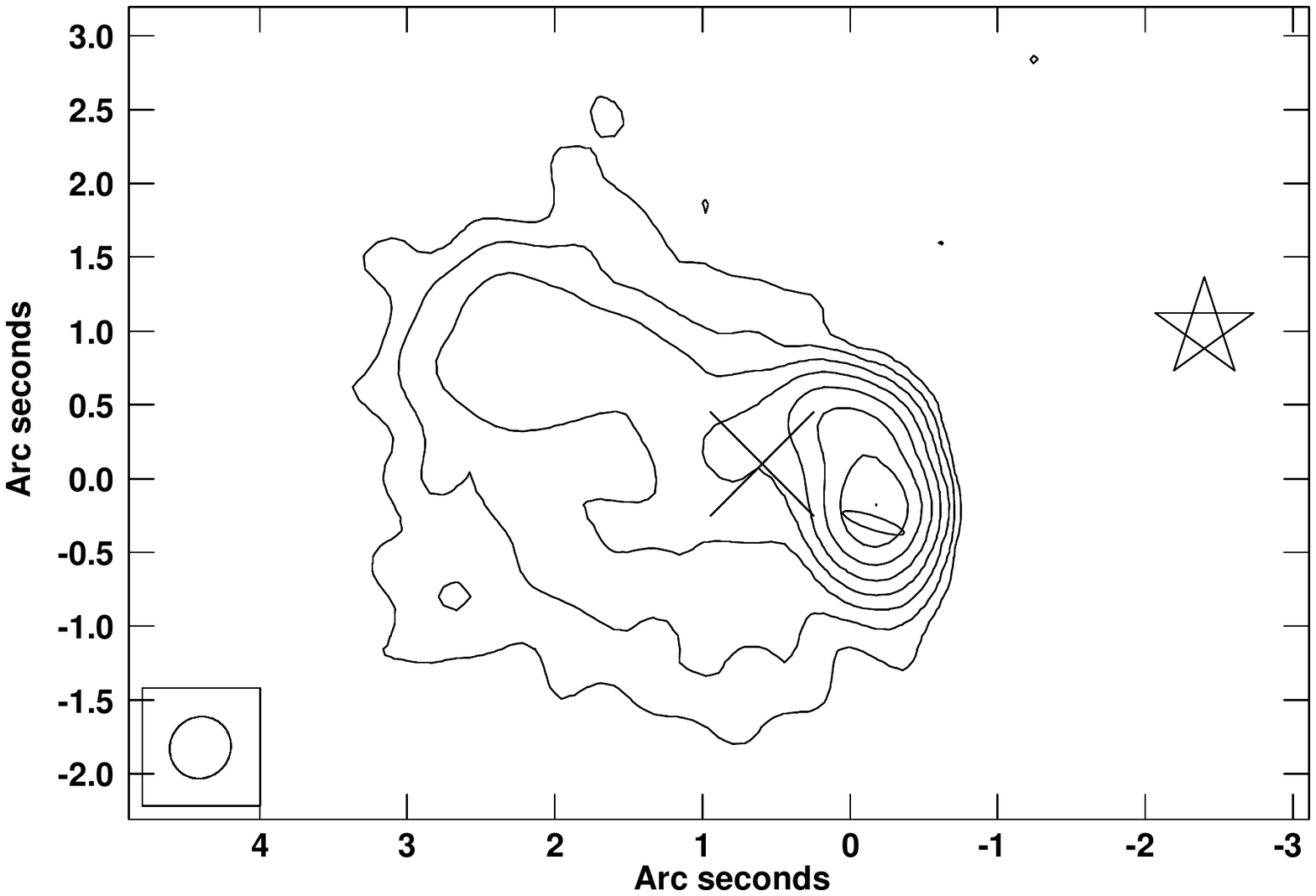}}
\caption{$4.86$-GHz VLA-A map of Complex B. Peak intensity is $78.7 \mathrm{\,mJy\,beam}^{-1}$, lowest contour levels at $3 \sigma$ rms noise level, 0.6 mJy/beam. Further contour levels increase by a factor of two. Beam size is $0\farcs43 \times 0\farcs41$ at a position angle of $-39\degr$ shown at the lower left part of the image. Cross indicates the pointing position of the EVN experiment EL034A, star indicates the pointing coordinates of the VLBA experiment BL189. Ellipse shows the position of the fitted elliptical Gaussian component to the EVN data. Its size is multiplied by 5 for clarity.}
\label{fig:B_VLAA_Cband}
\end{figure}

\begin{figure}
\centering
\resizebox{\hsize}{!}{\includegraphics[bb=13 8 530 445, clip]{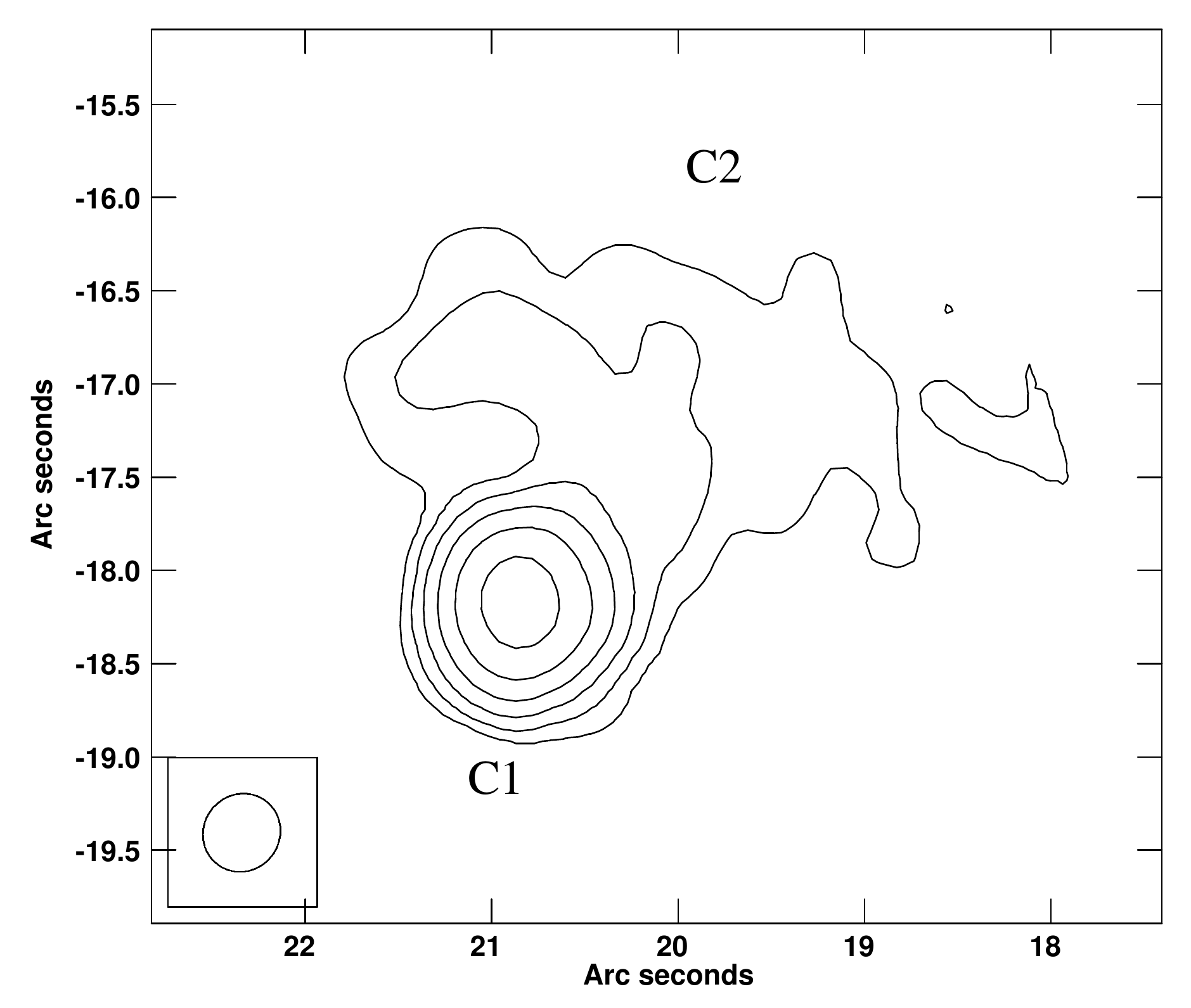}}
\caption{$4.86$-GHz VLA-A map of Complex C. Peak intensity is $27.8 \mathrm{\,mJy\,beam}^{-1}$, lowest contour levels at $3 \sigma$ rms noise level, $0.6 \mathrm{\,mJy\,beam}^{-1}$. Further contour levels increase by a factor of two. Beam size is $0\farcs43 \times 0\farcs41$ at a position angle of $-39\degr$ shown at the lower left part of the image.}
\label{fig:C_VLAA_Cband}
\end{figure}

\begin{figure}
\centering
\resizebox{\hsize}{!}{\includegraphics[bb=200 90 605 540, clip]{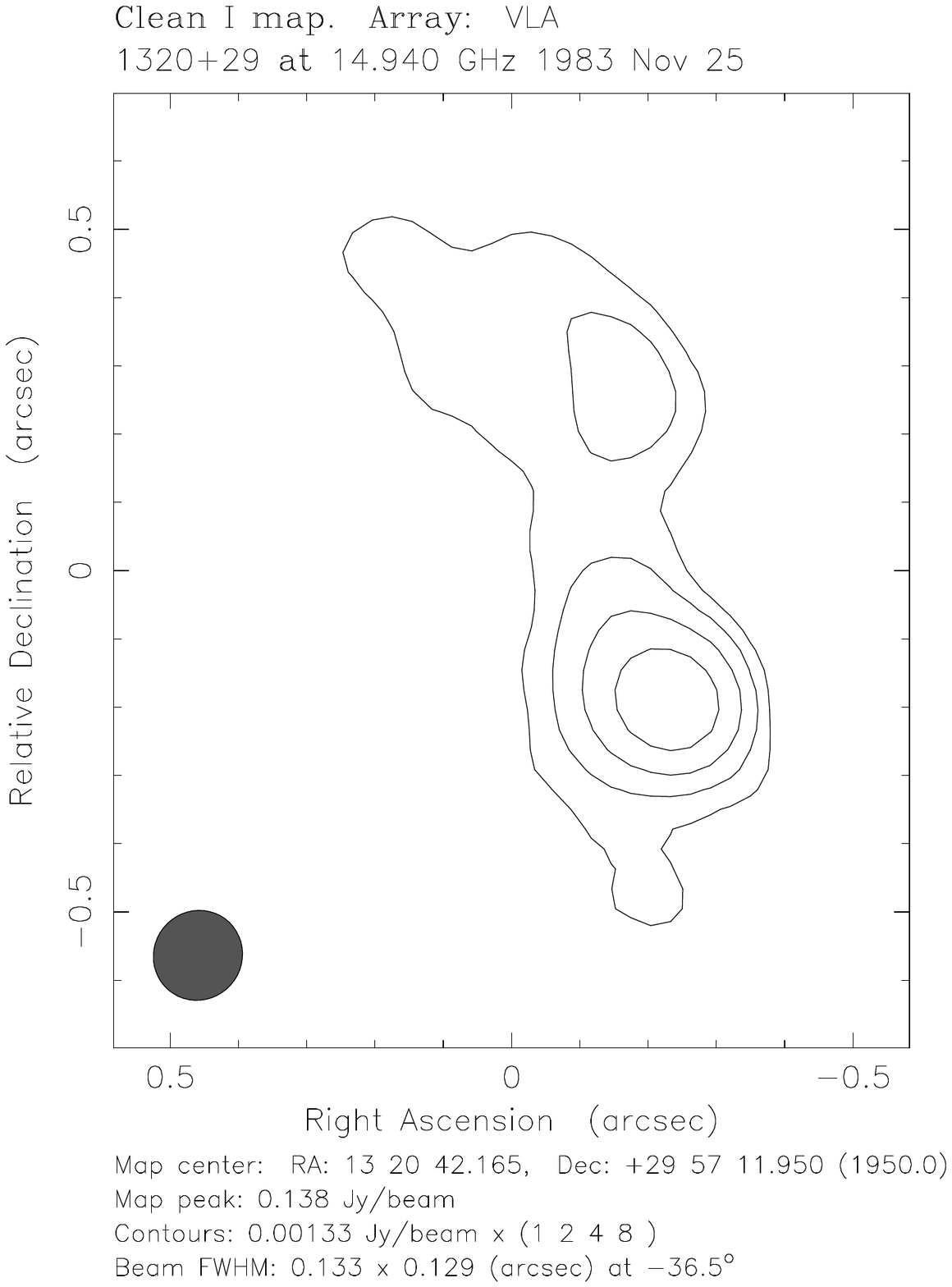}}
\caption{VLA-A map of Complex B at 14.94 GHz. Peak intensity is $19.2\mathrm{\,mJy\,beam}^{-1}$, lowest contour levels at $4 \sigma$ rms noise level, $1.3 \mathrm{\,mJy\,beam}^{-1}$. Further contour levels increase by a factor of two. The restoring beam is circular with a full-width half maximum (FWHM) size of $0\farcs13$ shown at the lower left part of the image. Coordinates are given relative to the pointing coordinate, right ascension $13^\textrm{h}23^\textrm{m}02\fs5370$ and declination $+29\degr 41' 33\farcs158$.} 
\label{fig:B_VLAA_Uband}
\end{figure}

\begin{table}
\caption{Results of modeling the brightness distribution of the VLA observations.}
\label{tab:VLA}
\centering
\begin{tabular}{ccccc}
\hline \hline
Frequency & ID & Flux density & R.A. & Dec. \\
(GHz) & & (mJy) & ($''$) & ($''$) \\
\hline 
1.49 & A1 & $376 \pm 20$ & $-21.6$ & $11.6$\\ 
 & A2 & $27 \pm 1$ & $-17.9$ & $10.8$ \\ 
 & B & $674 \pm 30$ & $0.0$ & $0.0$ \\
 & C1 & $168 \pm 8$ & $20.7$ & $-18.2$ \\
 & C2 & $80 \pm 4$ & $20.4$ & $-17.3$ \\ 
\hline 
4.86 & A1 & $239 \pm 10$ & $-21.6$ & $11.5$ \\
 & A2 & $6 \pm 0.5$ & $-17.8$ & $10.8$ \\
 & B & $219 \pm 11$ & $-0.2$ & $-0.2$ \\
 & C1 & $59 \pm 5$ & $20.9$ & $-18.2$ \\
 & C2 & $19 \pm 1$ & $20.3$ & $-17.1$ \\
\hline
14.94 & A1 & $(167 \pm 2)$ & $(-21.7)$ & $(11.6)$ \\
 & B & $49 \pm 3$ & $-0.2$ & $-0.2$ \\
 & C1 & $(14 \pm 3)$ & $(20.8)$ & $(-18.2)$ \\
\hline
\end{tabular}
\tablefoot{For details on the flux density see text. In cols. 4 and 5, relative coordinates are given for the brightest component in each region. The uncertainties of the coordinates are $0\farcs3$ at $1.49$\,GHz, and $0\farcs1$ at $4.86$ and $14.94$\,GHz. In the case of the $14.94$-GHz observation, the parentheses indicate that at the positions of A1 and C1 bandwidth smearing is significant, which could affect the results of the modelfit.}
\end{table}

\begin{figure}
\centering
\resizebox{\hsize}{!}{\includegraphics[bb=0 40 710 530, clip]{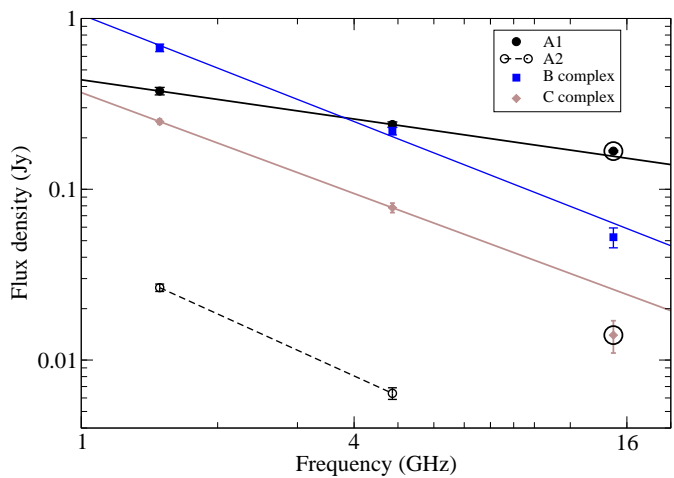}}
\caption{Radio spectra of the detected radio components in the VLA A-configuration observations. Lines represent the best-fit power law spectra, while symbols are flux density measurements from the modelfits. Filled black circles are for region A1, open black circles are for region A2, blue squares are for Complex B, filled brown diamonds are for Complex C. The two points marked by black circles are not used in the fitting.}
\label{fig:spectral_index}
\end{figure}

We were able to reproduce the total intensity maps published by \cite{Saikia1984} and \cite{Cornwell1986}. The $4.85$-GHz VLA A-configuration maps are displayed in Figs. \ref{fig:VLAA_Cband},  \ref{fig:A_VLAA_Cband}, \ref{fig:B_VLAA_Cband}, and \ref{fig:C_VLAA_Cband}. The coordinates given in the zoomed-in figures of the three radio complexes are relative to the pointing coordinate, right ascension $13^\textrm{h} 23^\textrm{m} 02\fs537$, and declination $+29\degr 41' 33\farcs16$.
Moreover, in VLA A-configuration data at $14.94$\,GHz where \cite{Cornwell1986} could only map Complex A, thanks to the modern data reduction software, we were able to image Complex B as well (Fig. \ref{fig:B_VLAA_Uband}). Additionally, at the position of Complex C, a faint feature at $4\sigma$ image noise level corresponding to $1.3$\,mJy\,beam$^{-1}$ could be tentatively detected at $14.94$\,GHz. 

At the highest frequency in A configuration, bandwidth smearing is significant at the positions of Complexes A and C, which are $\sim 25''$ and $\sim 28''$ away from the phase tracking center, respectively. The decrease of a point source response at these positions is $\sim 20$ per cent \citep{BW_smearing}. Correcting for this effect, Complex C has a peak brightness of $\sim 6.5$\,mJy\,beam$^{-1}$ at $14.94$\,GHz.

We used {\sc difmap} to model the brightness distributions of the three complexes with circular and elliptical Gaussian components at the three observing frequencies.
Complex A can be modeled by two circular Gaussians, a compact, bright ``core'' (A1), and a fainter feature (A2) at $\sim 3\farcs5$ toward east at $1.49$\,GHz. These two features can be seen as well at $4.86$\,GHz, however the higher resolution enabled to resolve the brightness distribution of A1 into three components, the brightest central, and two additional fainter features to the north and to the east. At $14.94$, only the brightest component A1 can be fitted with one circular Gaussian component. 

Complex B can be fitted by two circular Gaussian model components at $1.49$ and $14.94$\,GHz. At $4.85$\,GHz, several circular and elliptical Gaussian components were needed to reach the best fit to the visibilities. In Table \ref{tab:VLA}, we give the flux density sum of the fitted components, and the position of the brightest one. 

Complex C can be described by two circular Gaussian features at $1.49$ and $4.85$\,GHz (C1 and C2). At  $14.94$\,GHz, Complex C is close to the noise limit. However a stable modelfit result was achieved by adding a circular Gaussian component at the position of C1 region. 

We calculated spectral indices separately for the two features in Complex A, A1, and A2. In the case of the other two radio complexes, we give one spectral index for each, by summing up the flux densities of the Gaussian components of the model. The spectral index, $\alpha$ is defined as $S\sim \nu^\alpha$, where $S$ is the flux density and $\nu$ is the observing frequency. For Complex B, we were able to calculate the spectral indices using the results obtained at all three observing frequencies. In the case of A1 and Complex C, even though we were able to modelfit them at $14.94$\,GHz, we did not use their flux densities because the significant bandwidth smearing at their locations might influence the derived value. Only A1 has a flat radio spectrum, with $\alpha_\mathrm{A1}=-0.38 \pm 0.08$. The spectral indices of Complex B and Complex C agree within their uncertainties, $\alpha_\mathrm{B}=-1.0 \pm 0.1$, and $\alpha_\mathrm{C}=-0.98 \pm 0.08$.
The $1.49$--$4.86$\,GHz spectral index of A2 is also similar, $\alpha_\mathrm{A2}=-1.2 \pm 0.03$. The radio spectra are shown in Fig. \ref{fig:spectral_index}. We also plot here the two flux density measurements not used for the spectral index calculation, marked by large black circles.

\subsection{Very long baseline interferometric observations of Complex B}

\begin{figure*}
\begin{minipage}{0.5\textwidth}
\centering
\includegraphics[width=\columnwidth, bb=0 0 720 515, clip]{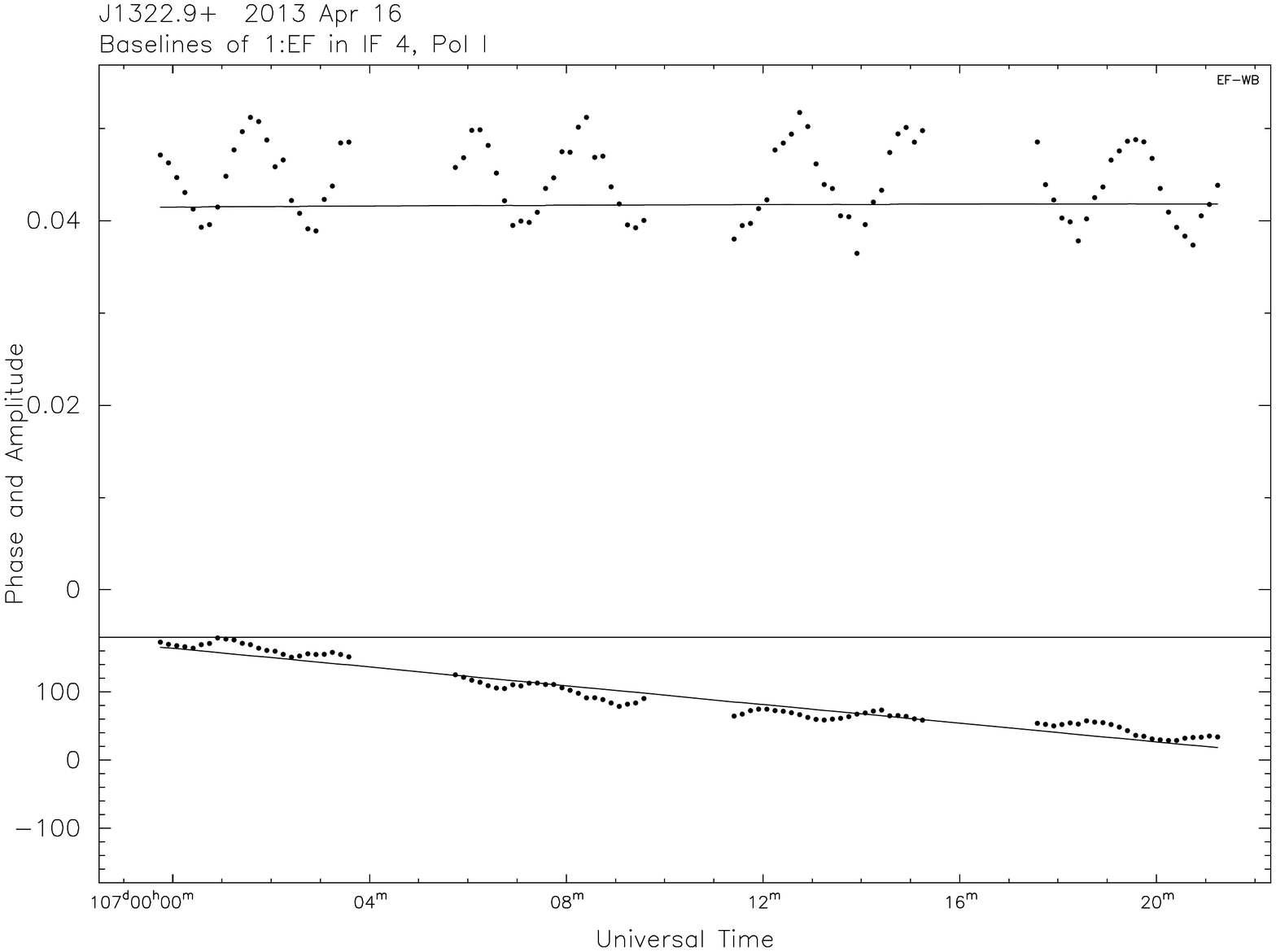}
\end{minipage}
\hfill
\begin{minipage}{0.5\textwidth}
\includegraphics[width=\columnwidth, bb= 0 0 720 515, clip]{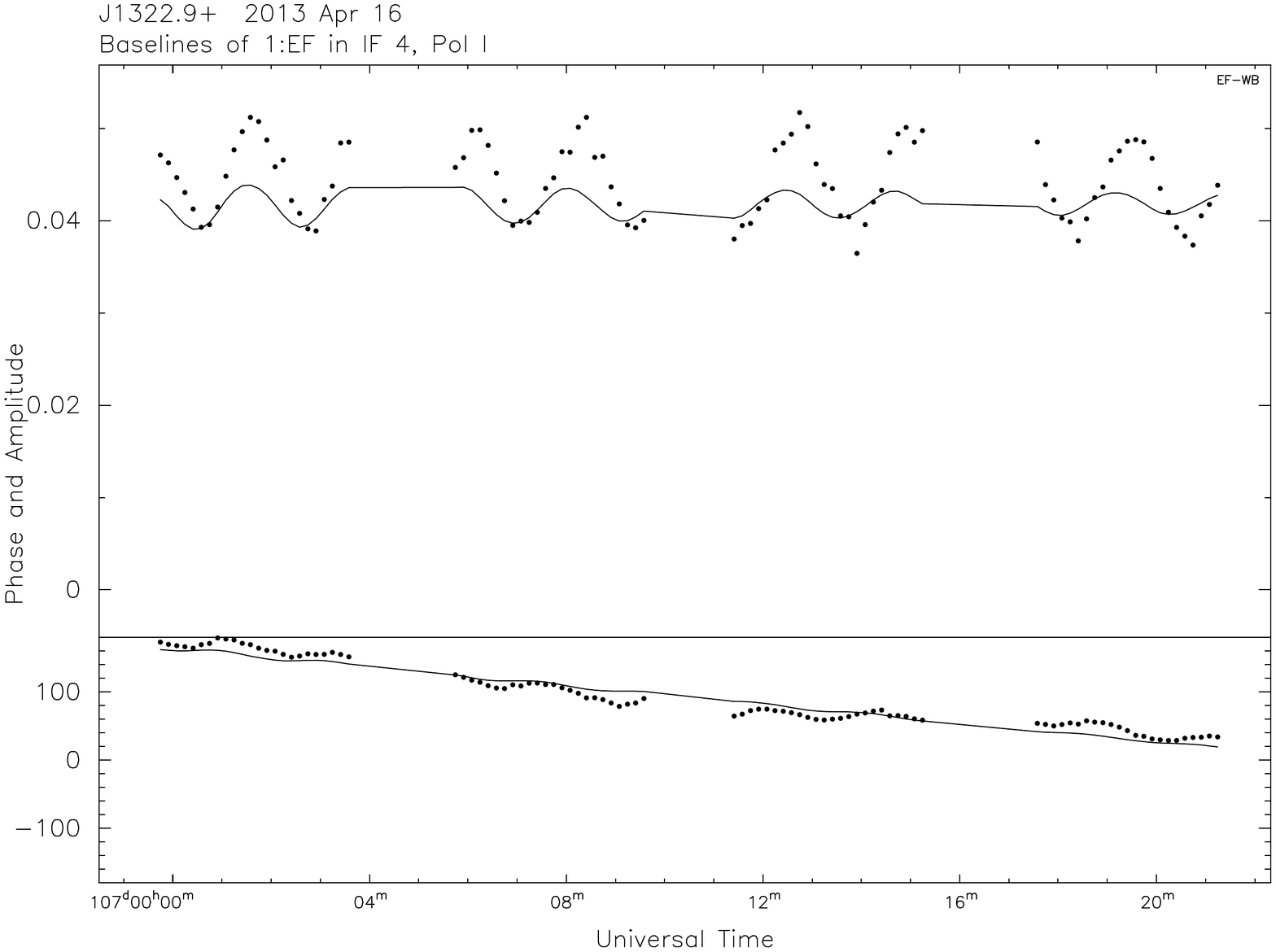}
\end{minipage}
\caption{Visibility amplitudes in Jy (top) and phases in degrees (bottom) versus time from the EVN experiment EL034A on the Effelsberg--WSRT baseline for IF 4. Measured data are shown as points, lines represent the visibility model. {\it Left:} The model contains one elliptical Gaussian component. {\it Right:} The model also contains an additional elliptical Gaussian component around the position of Complex A.}
\label{fig:radplot}
\end{figure*}

An extended radio source can be detected at $1.6$\,GHz in the EVN observation which targeted Complex B. Using the {\sc difmap} package to fit the visibilities, this feature can be best described by an elliptical Gaussian brightness distribution at a position a few hundred mas southwest from the pointing position, close to the peak of Complex B in the VLA images, as shown by the ellipse in Fig. \ref{fig:B_VLAA_Cband}. Its flux density is $44.1\pm 9.3$\,mJy, the full width at half maximum (FWHM) major axis size is $87 \pm 18$\,mas, and its axial ratio is $0.20\pm 0.07$. This single elliptical Gaussian can adequately describe most of the visibility measurements, however there is a short timescale periodical variation in the visibilities at the Effelsberg--WSRT baseline, which cannot be fitted with this simple model (see left hand side of Fig. \ref{fig:radplot}). Therefore, we included an additional elliptical component to describe the brightness distribution of the bright radio source related to Complex A. This model provided a better fit to the visbilities on the Effelsberg--WSRT baseline (see right hand side of Fig. \ref{fig:radplot}). Even though Complex A is located well outside of the undistorted field of view of the array, its contribution to the visbilities can be clearly detected at the shortest and most sensitive baseline. Because of the substantial smearing, the flux density and size of the detected radio feature in Complex A cannot be firmly deduced from these data.

The VLBA experiment BL189 targeted a slightly different position than the EVN observation (Fig. \ref{fig:B_VLAA_Cband}). In this experiment, the found fringes are from the bright radio-emitting feature in Complex A, at $\sim 21''$ away from the pointing position. Fringes could not be detected for Complex B.

\subsection{Very Long Baseline Array observations of Complex A}

\begin{figure*}
\begin{minipage}{0.47\textwidth}
\centering
\includegraphics[width=\columnwidth, bb=0 0 630 670, clip]{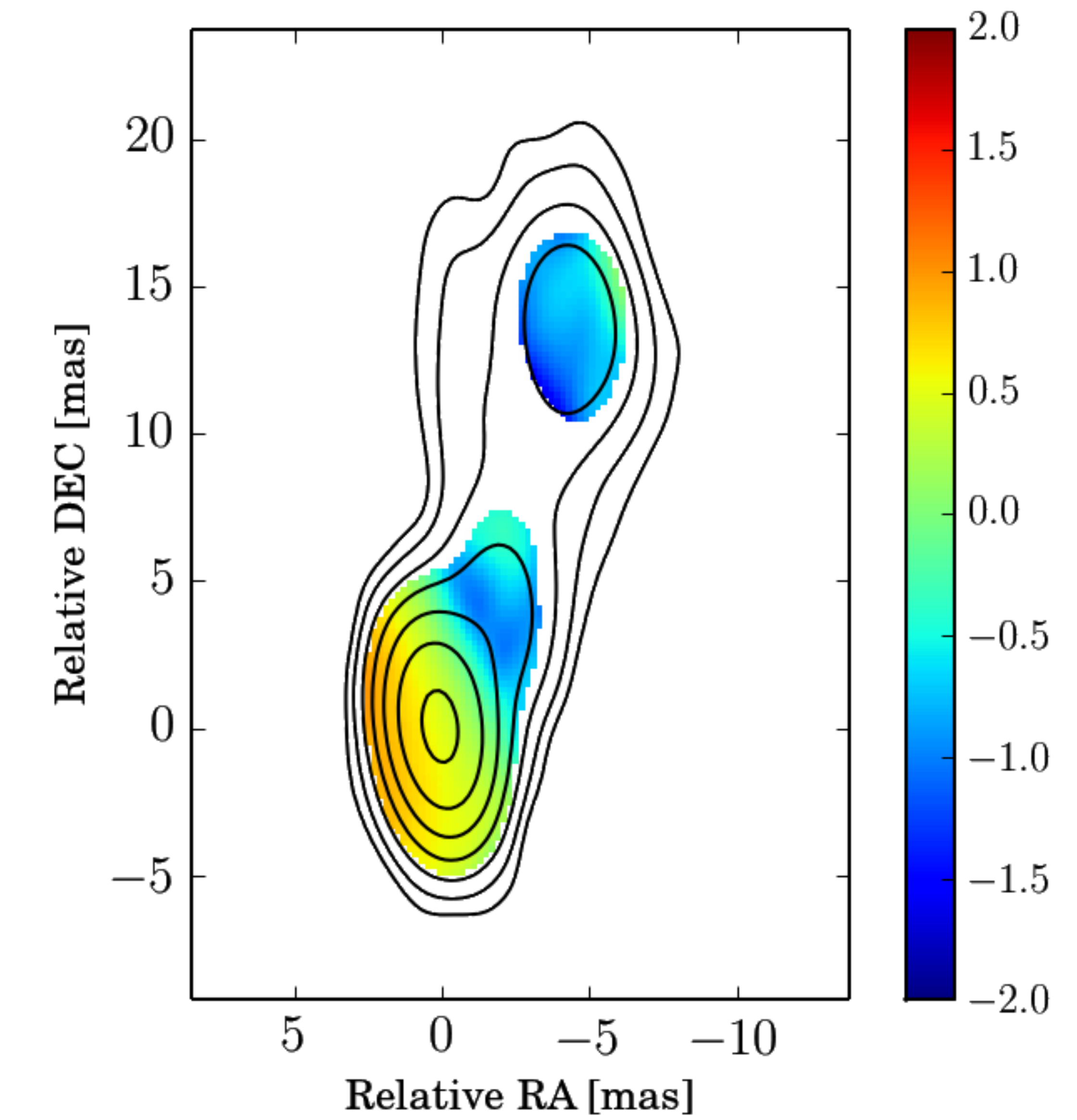}
\caption{VLBA image of Complex A. The color scale represents the spectral index distribution between $4.3$ and $7.6$\,GHz, overlaid on the contours showing the 4.3\,GHz map of the source. Peak intensity is $102 \textrm{\,mJy\,beam}^{-1}$, the lowest contour level is at $1.35 \textrm{\,mJy\,beam}^{-1}$ corresponding to $3\sigma$ image noise level. The beam size is $4.9 \textrm{\,mas} \times 2.3 \textrm{\,mas}$ at a position angle of $10\degr$. The image was created using the {\sc vimap} program \citep{vimap}.}
\label{fig:VLBA_spectral_index}
\end{minipage}
\hfill
\begin{minipage}{0.47\textwidth}
\centering
\includegraphics[width=\columnwidth,bb=200 105 640 540, clip]{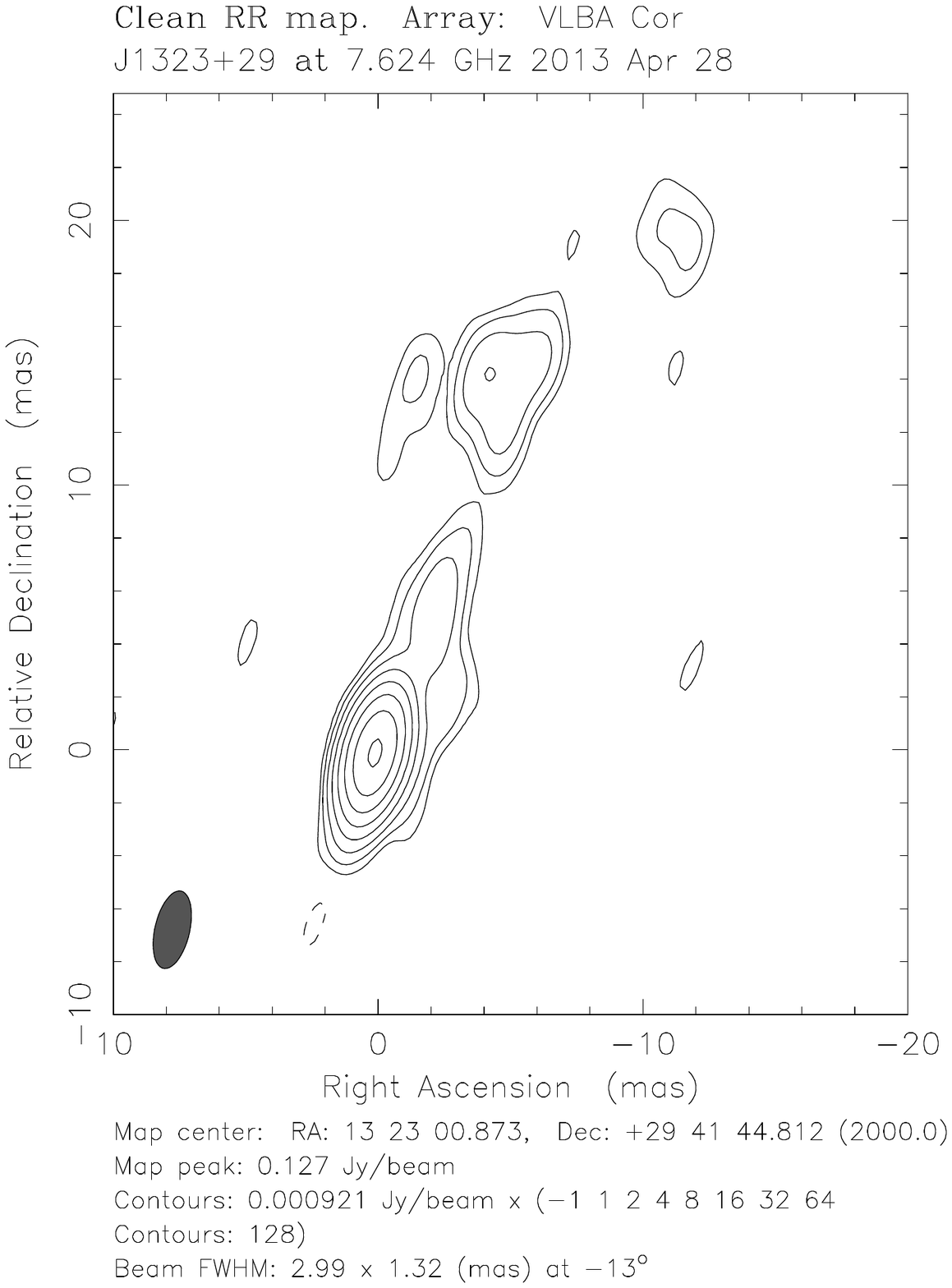}
\caption{$7.6$-GHz VLBA image of Complex A. Peak intensity is $127 \textrm{\,mJy\,beam}^{-1}$, lowest positive contour is at $5\sigma$ image noise level, $0.9 \textrm{\,mJy\,beam}^{-1}$. Further contours increase with a factor of two. The restoring beam is $3 \textrm{\,mas} \times 1.3\textrm{\,mas}$ at a position angle of $-13\degr$, and it is shown in the lower left corner of the image.}
\label{fig:VLBA_X}
\end{minipage}
\end{figure*}

The VLBA images of Complex A\footnote{Listed under the name J1323$+$2941 on the astrogeo.org website} show an elongated core--jet structure oriented to north-northwest.
At $4.3$\,GHz, the jet-like feature connecting the southern brightest core component and the possible blob in the jet is clearly detected (Fig. \ref{fig:VLBA_spectral_index}). At $7.6$\,GHz, the jet and this northern feature are starting to get resolved out (Fig. \ref{fig:VLBA_X}). 

We used the {\sc vimap} program created by \cite{vimap} to produce the spectral index map between these two observing frequencies. The $7.6$-GHz image was created with the same pixel size and restoring beam size as was used for the $4.3$-GHz map. After excluding the core region in both images, {\sc vimap} calculates the cross-correlation product to determine the shift between the two images. In our case, because of the relatively close frequencies, there was no need to shift the data. The resulting spectral index map is shown in Fig. \ref{fig:VLBA_spectral_index}, overlaid on the $4.3$-GHz VLBA contour map. The spectral index distribution confirms the core--jet source structure. Indeed the brightest southern feature has flat spectral index, $\sim 0.5$, as expected from the VLBI core of beamed AGN, while the northern region has a steeper spectrum with a spectral index of $\sim -1.0$ typical of blazar jets \citep[e.g.,][]{hovatta_spectralindex}.

We used the {\sc difmap} software to fit the brightness distribution of the source with Gaussian components. At both frequencies, our best-fit model to the visibility data contained four Gaussian components. 
At $4.3$\,GHz, the core was fitted by the brightest and most compact circular component. Its flux density and FWHM size are $S_\mathrm{core}^{4.3\mathrm{\,GHz}}=(92.7 \pm 8)$\,mJy and $\theta_\mathrm{core}^{4.3\mathrm{\,GHz}}=(0.58 \pm0.03)$\,mas, respectively. 
At $7.6$\,GHz, the core was best fitted by an elliptical Gaussian with the following parameters: flux density $S_\mathrm{core}^{7.6\mathrm{\,GHz}}=(141 \pm 10)$\,mJy, major axis FWHM size $\theta_\mathrm{core}^{7.6\mathrm{\,GHz}}=(0.72 \pm 0.04)$\,mas and axial ratio $a=0.6 \pm 0.1$. 

The brightness temperature can be calculated as
\begin{equation}
T_\textrm{B}=1.22 \times 10^{12} \left(1+z\right) \frac{S}{a\theta^2\nu^2}\,\mathrm{K} 
,\end{equation}
where $z$ is the redshift, $S$ is the flux density in Jy, $\nu$ is the observing frequency in GHz, $\theta$ is the major axis size of a Gaussian component in mas, and $a$ is the axial ratio. Using the values derived from the $7.6$-GHz observation, the brightness temperature is $T_\mathrm{B}=9.6 \times 10^9 (1+z)$\,K. The redshift of the optical quasar coincident with the radio source is $z=1.142$ (for details see Sect. \ref{sec:disc}), thus the brightness temperature is $T_\mathrm{B}=(2.1 \pm 0.4) \times 10^{10}$\,K. This high value clearly indicates AGN origin of the radio emission. On the other hand, it is below the commonly used equipartition brightness temperature limit of $\sim 5 \times 10^{10}$\,K \citep{Readhead}, and even below $3\times 10^{10}$\,K, the characteristic brightness temperature of pc-scale radio cores found by \cite{Homan_TB}. Thus, it does not seem to imply Doppler boosting. However, the fact that elliptical Gaussian provided the best fit to the core brightness distribution may indicate that the observation could not resolve the blended VLBI core and the closest jet component. This is also supported by the fact that the major axis of the fitted elliptical component is in the jet direction, at $\sim -25\degr$. Therefore the size can be regarded as an upper limit and the brightness temperature as a lower limit only.

\section{Discussion} \label{sec:disc}

\subsection{The origin of $\gamma$-ray emission}

Already \cite{Cornwell1986} discussed the possible nature of 4C$+$29.48 in the light of VLA observations. \cite{Cornwell1986} pointed out that since there is no evidence for any physical connection between the three radio sources (Complex A, B, and C), they can be completely unrelated objects. They also found that Complex A can be identified with an optical quasar. They proposed two possibilities for Complex B and C: either they are unrelated radio sources, then B can be a head-tail radio galaxy, or they can be the two lobes of a radio galaxy too faint in the optical to be detected at that time. 

More recently, \cite{VLBI_imaging_Morgan} could strengthen the quasar identification of Complex A, since they were able to detect compact pc-scale radio emission there at $8.4$\,GHz. They conclude that their non-detection of compact radio source in Complex B supports the suggestion of \cite{Cornwell1986} that this source may be a head-tail radio galaxy. With respect to Complex C, the non-detection of compact radio emission is in disagreement with its classification of being an edge-brightened hot spot.

We can confirm the results of \cite{Cornwell1986} that Complex A is a separate source from Complexes B and C. Even using the more up-to-date data analysis software, we were not able to detect any faint radio features between the three complexes at any of the VLA observations down to a $3\sigma$ image noise level of $2.6\mathrm{\,mJy\,beam}^{-1}$ in the VLA C-configuration data, $2.0\mathrm{\,mJy\,beam}^{-1}$ in the VLA B-configuration data, and $0.6\mathrm{\,mJy\,beam}^{-1}$ in the highest resolution VLA A-configuration data. The derived spectral indices of the brightest features of the three complexes (Fig. \ref{fig:spectral_index}) agree with those given by \cite{Cornwell1986}, even for Complex B for which \cite{Cornwell1986} could give only two-point spectral indices, but we could detect it also at the highest frequency. Only A1 has a flat radio spectrum, all other radio features have steep spectra, indicating their probable AGN lobe-related nature. The faint extension connecting A1 and A2 in the $1.49$-GHz and $4.86$-GHz images may indicate that they are related, A2 being a brighter jet feature or hotspot in the jet of the quasar.

Similarly to \cite{VLBI_imaging_Morgan} we could not detect compact radio emission in Complex B in the archival VLBI data. However, an extended radio-emitting feature can be seen at the 1.6-GHz EVN observation at this position. In Complex A, there is a high brightness temperature, flat-spectrum VLBI core with a steep-spectrum jet pointing north-northwest. The steep-spectrum radio feature (A2) detected in the VLA observations at $\sim 3\farcs8$ from A1 is outside the undistorted field of view of the VLBA observations. Its flux density measured at $4.86$\,GHz with the VLA, $(6.0 \pm 0.5)$\,mJy, would not allow for a detection at such distance from the pointing center. 

To better understand 4C$+$29.48 and its $\gamma$-ray emission, we complemented the radio observations with data taken at other wavebands. In the Sloan Digital Sky Survey 12th data release \citep[SDSS DR12,][]{sdss_dr12}\footnote{http://skyserver.sdss.org/dr12/en/home.aspx}, there is one optical source, SDSS\,J132300.86$+$294144.8, with a position coincident with Complex A (hereafter SDSS-A). It is classified as quasar, its measured redshift is $1.142$.\footnote{The VERITAS non-detection of the source in TeV $\gamma$-rays \citep{VERITAS_non} is not surprising given that the currently known highest redshift TeV sources are at $z\sim1$ \citep{highz_TeV}.} This source is also listed in the first data release of the {\it Gaia} mission \citep{GaiaDR1, GaiaMission} with more accurate coordinates. The right ascension is $13^\textrm{h} 23^\textrm{m}00\fs87334$ with an uncertainty of $0.452$\,mas, and the declination is$+29\degr 41\arcmin 44\farcs8167$ with an uncertainty of $0.86$\,mas.

There is no coincident optical object in the SDSS at the positions of Complexes B or C. However, there is a faint optical source, SDSS\,J132303.03$+$294126.4 (hereafter SDSS-Z) between Complexes B and C. Its position is shown in Fig. \ref{fig:VLAA_Cband}. According to SDSS DR12, the photometry of this faint object is unreliable. (Persumably because of its faintness it was not detected by {\it Gaia}.)

In the infrared catalog based upon the measurements of the Wide-field Infrared Survey Explorer \citep[{\it WISE},][]{wise} \citep[AllWISE, ][]{AllWISE} we looked for counterparts of the sources. WISE J132300.86$+$294144.7 (hereafter WISE-A) is positionally coincident with Complex A, while WISE J132303.04$+$294126.7 (hereafter WISE-Z) is located between Complexes B and C, and positionally coincident with the faint optical source, SDSS-Z found there. WISE-A was detected in three out of the four WISE bands, at $3.35\mu$m, $4.6\mu$m, and $11.6\mu$m with a signal-to-noise ratio (S/N) larger than five. In the longest-wavelength band, at $22 \mu$m, only an upper limit can be given for its brightness. WISE-Z was detected in all four {\it WISE} bands with an S/N larger than five. 

\begin{figure}
\centering
\resizebox{\hsize}{!}{\includegraphics[bb=30 40 710 530, clip]{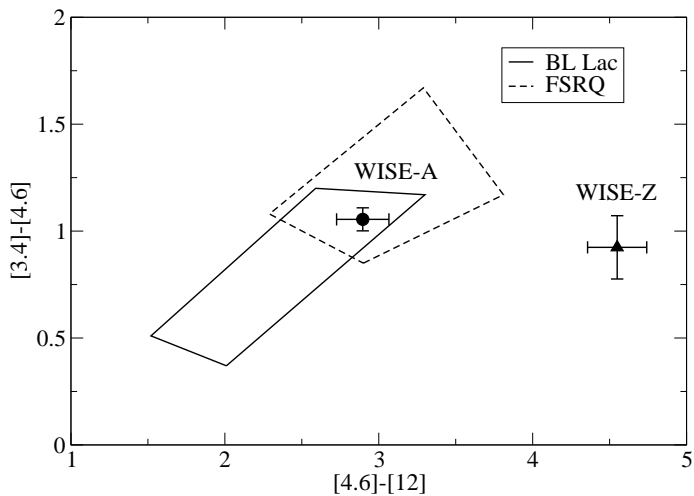}}
\caption{WISE Gamma-ray Strip following \cite{WGS_1} and \cite{WGS} are shown for BL Lac objects (solid line) and Flat Spectrum Radio Quasars (dashed line) on a color--color diagram using the three shorter wavelength bands of {\it WISE}. The circle represents the infrared colors of WISE-A, the triangle represents the infrared colors WISE-Z.}
\label{fig:wise}
\end{figure}

Based upon the {\it WISE} observations in the three shorter-wavelength bands, \cite{WGS_1} defined the WISE Gamma-ray Strip \citep[WGS, see also][]{WGS}, the area in the infrared color--color diagram where the {\it Fermi}-detected blazars are located. In Fig. \ref{fig:wise}, we show the position of the two {\it WISE} sources with respect to the two subregions of the WGS, the area occupied by BL Lac objects and by FSRQs. The infrared colors of WISE-A place it within the WGS. However, WISE-Z is clearly out of this region, its infrared colors are incompatible with those sources belonging to the WGS. According to the color--color diagram of \cite{wise} and \cite{Massaro_Fermi_review}, the mid-infrared characteristics of WISE-Z place it closer to the region where the luminous infrared galaxies, and ultra-luminous infrared galaxies are located. In Table \ref{tab:coord}, we summarize the above discussed sources, their coordinates and possible counterparts.

There is one X-ray source, 2RXS J132302.9$+$294115, in the $0.1-2.4$\,keV energy range listed in the Second ROSAT all-sky survey source catalog \citep{rosat_Boller} close to region occupied by 4C$+$29.48. Its location with respect to the three radio complexes is shown in Fig. \ref{fig:FIRST_Fermi}. \cite{Fermi3_AGNcat} claim that the probability of the X-ray--$\gamma$-ray association is below their threshold of $0.8$. The position of the X-ray source might indicate its possible relation to the optical and infrared source between Complexes B and C (i.e., SDSS-Z, WISE-Z).

In Fig. \ref{fig:FIRST_Fermi}, we overplot the $95$\,per cent confidence  radius of the {\it Fermi} detection in the FIRST image of 4C$+$29.48. It is clear that Complex A is closer to the center of the confidence region than Complex B. Also the combined radio, optical, and infrared data all agree with Complex A being a blazar, which make up $98$ per cent of the {\it Fermi}-detected extragalactic gamma-ray sources \citep[e.g.,][]{Fermi3_AGNcat}. 

\begin{figure}
\centering
\resizebox{\hsize}{!}{\includegraphics[bb=20 5 420 310, clip]{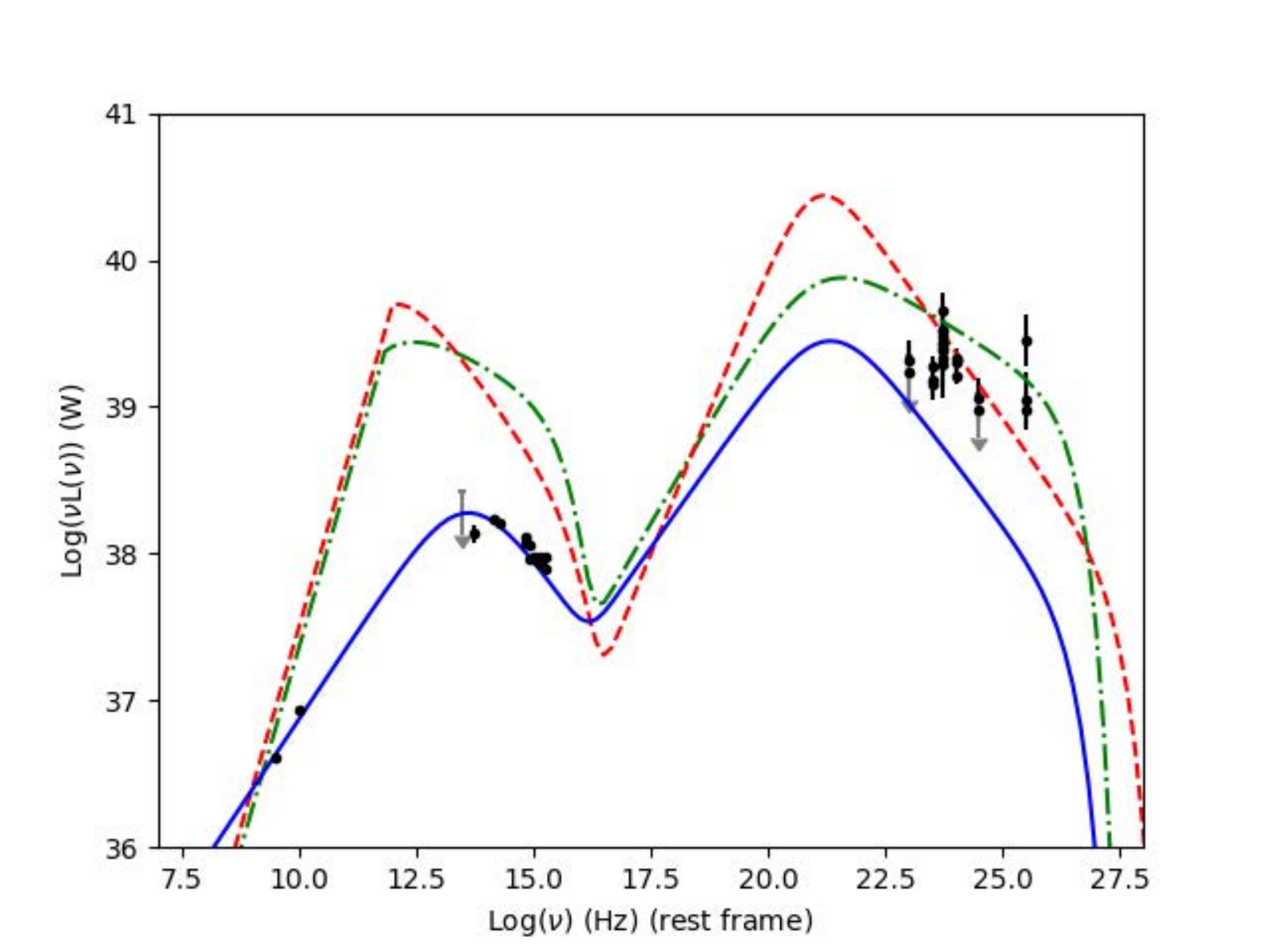}}
\caption{SED of the source in Complex A (calculated in the rest frame of the source at $z=1.142$). Black dots represent radio (VLA A-configuration), infrared (WISE), optical (SDSS), and $\gamma$-ray ({\it Fermi}) observations. Gray arrows indicate upper limits. Dashed red, and dotted-dashed green lines are the SED derived by \cite{Fermi_blazarseq} for the {\it Fermi}-detected FSRQ and BL Lac objects, respectively, in the luminosity bin of $10^{40}-10^{41}$\,W. Solid blue line is obtained via fitting the synchrotron peak and keeping the inverse Compton peak at the frequency derived by \cite{Fermi_blazarseq} (for further details see text).}
\label{fig:SED}
\end{figure}

\cite{Fermi_blazarseq} described the average SED of the {\it Fermi}-detected BL Lac objects and FSRQs. In Fig. \ref{fig:SED}, we plot the  radio (VLA A-configuration flux density of A1), the infrared (WISE-A) and the optical data (SDSS-A) of Complex A together with the $\gamma$-ray data measured by {\it Fermi} LAT. The latter were obtained through the portal of the Space Science Data Center (SSDC) SED builder tool\footnote{\url{http://tools.asdc.asi.it/SED/}} using the following {\it Fermi} catalogs of \cite{SED-1FGL}, \cite{SED-2FGL}, and \cite{FermiLAT3}. We overplot the curves from \cite{Fermi_blazarseq} describing the average BL Lac object and FSRQ obtained for the luminosity bin $10^{40}-10^{41}$\,W. While these curves are close to the $\gamma$-ray measurements, they vastly overestimate the low-frequency data points. 

We fitted a third-order polynomial to the radio, infrared and optical data to estimate the position and luminosity of the synchrotron peak, $3.6 \cdot 10^{13}$\,Hz, and $5 \cdot 10^{24}$\,W, respectively. \cite{Fermi_blazarseq} fixed the radio part of their SED to a power-law with an exponent of $-0.1$. In the case of Complex A we found that using the derived values for the synchrotron peak, a single curve can provide an adequate description of the data from the radio to the optical (blue solid line in Fig. \ref{fig:SED}). The equation of the curve is given in Eq. (3) of \cite{Fermi_blazarseq}, the parameters are $\nu_S = 3.6 \cdot 10^{13}$\,Hz (the frequency of the synchrotron peak), $\alpha_1=0.52$, $\alpha_2 = 1.42$, and $\nu_\mathrm{cut,S} = 10^{16}$\,Hz.

The inverse Compton peak is much less constrained. The largest $\gamma$-ray flux was reported in the light curve published in the second {\it Fermi} LAT catalog \citep[2FGL]{SED-2FGL}, where the photon flux over the whole energy range (between 100 MeV and 100 GeV) is given. These measurements are shown at $10$\,GeV in Fig. \ref{fig:SED}. Assuming that the inverse-Compton peak is at the lowest {\it Fermi} band at $200$ MeV, the largest $\gamma$-ray flux measured there is $2 \cdot 10^{23}$\,W, resulting in a Compton dominance parameter (the ratio of $\nu L_\nu$ at the inverse Compton peak to the value at the synchrotron peak) of $11$. On the other hand, if we try to describe the high-energy part of the SED together with the curve derived for the synchrotron peak with high-energy parameters in the same range that of \cite{Fermi_blazarseq}, thus keeping the inverse Compton peak at $\sim 10^{21}$\,Hz, a Compton dominance parameter of $15$, which was the largest obtained by \cite{Fermi_blazarseq}, would still underestimate the $\gamma$-ray measurements, as shown by the solid blue curve in Fig. \ref{fig:SED}.

However, the measurements conducted in the different bands are not simultaneous, thus source variability may significantly influence the shape of the SED. According to the $\gamma$-ray light curve of \cite{SED-2FGL}, and also by comparing the values from the different {\it Fermi} catalogs, the source is highly variable. Therefore a reliable SED can only be constructed from simultaneous measurements. Additionally, observation of correlated variability in the optical and $\gamma$-ray bands would provide unequivocal evidence for the identification of the optical quasar at $z=1.142$ and the radio source in Complex A with the $\it Fermi$-detected $\gamma$-ray emission.

\begin{table*}
\caption{Names, most accurate coordinates, and counterparts of the sources discussed. In footnotes we give the references for the coordinates.}
\label{tab:coord}
\centering
\begin{tabular}{c|cc|ccc}
\hline \hline
 ID & \multicolumn{2}{c}{Best radio position} & \multicolumn{3}{|c}{Name of the counterpart}  \\
 & Right ascension & Declination & SDSS & {\it WISE} & {\it Fermi} \\
\hline
A & $13^\textrm{h} 23^\textrm{m}00\fs87350$ & $+29\degr 41\arcmin 44\farcs812$\tablefootmark{a} & J132300.86$+$294144.8 & J132300.86$+$294144.7 & 3FGL\,J1323.0$+$2942 \\
B & $13^\textrm{h} 23^\textrm{m}02\fs5191$ & $+29\degr 41\arcmin 32\farcs967$\tablefootmark{b}& -- & -- & -- \\
C & $13^\textrm{h} 23^\textrm{m}04\fs140$ & $+29\degr 41\arcmin 15\farcs03$\tablefootmark{c}& -- & -- & -- \\
Z & -- & -- & J132303.03$+$294126.4 & J132303.04$+$294126.7 & -- \\ 
\hline
\end{tabular}
\tablefoot{\tablefoottext{a}{$7.6$-GHz VLBA observations}\tablefoottext{b}{$14.94$-GHz VLA A-configuration observation  \citep{Cornwell1986}}\tablefoottext{c}{$4.86$-GHz VLA A-configuration observation \citep{Cornwell1986}}}
\end{table*}

\subsection{One object, two objects, or three?}

The nature of the radio sources of Complexes B and C is however still uncertain. In the following we discuss possible scenarios of the connections of the three radio complexes. 

In principle all of them can be related to one source. In that case, Complexes B and C can be jet features or hot spots in the jet of the blazar residing in Complex A. The angular separations of Complexes B and C from the blazar are $24\farcs5$ and $51\farcs2$, corresponding to projected linear sizes of $\sim 202$\,kpc and $\sim 422$\,kpc, respectively at the redshift of Complex A. Assuming that they are at $z=1.142$, their $1.49$-GHz radio powers are $P_\mathrm{B}=4.9 \times 10^{27} \mathrm{\,W\,Hz}^{-1}$, $P_\mathrm{C1}= 1.2\times 10^{27} \mathrm{\,W\,Hz}^{-1}$, and $P_\mathrm{C2}= 6.8\times 10^{26} \mathrm{\,W\,Hz}^{-1}$. These values alone do not contradict the proposed picture, for example, \cite{largescale_MOJAVE} studied the kpc-scale radio morphologies of a flux-limited sample of blazars and found radio powers as high as $10^{28} \mathrm{\,W\,Hz}^{-1}$. In this scenario one would expect to see similar jet-related features on the other side of Complex A, however there are no such radio- emitting features present there as can be seen for example, in the FIRST image in Fig. \ref{fig:FIRST_Fermi}. Additionally, the morphology of Complex B is hard to reconcile with a south-east directed jet, it may indicate interaction with the surrounding medium. 

Another possibility is that the three radio complexes belong to two different sources: Complex A is a blazar, and Complexes B and C are part of another object only in chance alignment with the blazar. Complexes B and C can be the two radio lobes of a radio galaxy as suggested already by \cite{Cornwell1986}. In that case the faint optical source, SDSS-Z (and the associated infrared source WISE-Z) between them could be identified with the host galaxy. In that scenario, the lobe located closer to the observer would be the brighter radio feature, Complex B. This is also supported by the fact that Complex B is more polarized than Complex C \citep{Cornwell1986}. In the usual picture of expanding radio galaxies, the advancing lobe (i.e., closer to the observer) is seen further away from the host galaxy in projection based on light travel time arguments \citep{arm-length-ratio}; the arm-length ratio of the bright to faint lobes are larger than one. However, Complex B, the brighter of the lobes is closer to the assumed host galaxy, the arm length ratio of Complex B to Complex C is $\sim 0.5$. This could mean that instead of projection effects, the distance to the host galaxy and the brightness of the radio lobes are mostly influenced by the surrounding medium, which is significantly asymmetric on the two sides of the host galaxy.

Alternatively, Complexes B and C could also be gravitationally lensed images of the same background source. Their radio spectral indices agree within the uncertainties, supporting this idea. If the two radio sources are indeed images of one background source, the image separation would be comparable to the the largest one reported so far, which is $22\farcs5$, in the case of SDSS\,J102913.94$+$262317.9 \citep{largest_sep_gravlens}. Such large image separation requires a massive lensing object - a cluster of galaxies \citep{Inada_nature}. In the case of SDSS\,J102913.94$+$262317.9, the possible lensing cluster of galaxies could be detected in optical wavelengths. In our case, however, there is no indication of a cluster near the radio sources in the SDSS.

According to \cite{Cornwell1986}, the probability of an unrelated source with $1.49$-GHz flux density similar to those of Complexes C or A at a distance of $\sim 25''$ from Complex B is $\sim 5 \times 10^{-5}$. If the radio features belong to different galaxies their proximity can be explained if their hosts reside in a galaxy cluster as suggested by \cite{Cornwell1986}. However, according to the SDSS images, there is no indication of the existence of a galaxy cluster in this region. 

\section{Summary} \label{sec:sum}

The $\gamma$-ray source 3FGL\,J1323.0+2942 is associated with the radio source 4C$+$29.48, and classified as a blazar of unknown type \citep{FermiLAT3}. Its elongated radio structure seen in the NVSS map is resolved into three bright radio features, Complexes A, B, and C, in the better-resolution FIRST image (Fig. \ref{fig:FIRST_Fermi}). VLBI observation of \cite{VLBA_incorrect_pointing_Lico} pointing close to the middle source, Complex B, was not able to detect compact radio emission. The wide-field imaging of \cite{VLBI_imaging_Morgan} showed that only Complex A includes pc-scale compact VLBI feature.

We analyzed archival, not yet published EVN and VLBA observations and re-analyzed VLA observations of \cite{Cornwell1986} and \cite{Saikia1984}. These revealed that Complex A is a blazar with a north-northwest jet structure at pc-scales, which is most probably bent by $\sim 90\degr$ to create the east-bound jet feature seen at kpc scales. Complexes C and B consist of features with steep radio spectra, they are more reminiscent of lobe-like radio emission.

The accurate radio coordinates of the blazar in Complex A are right ascension  $13^\textrm{h} 23^\textrm{m}00\fs87350$ and declination $+29\degr 41\arcmin 44\farcs812$. It has an optical counterpart in the SDSS DR12 at a redshift of 1.142, and an infrared counterpart in the AllWISE catalog. The infrared colors of the WISE source place it in an area of the WISE three-color color--color diagram where most of the $\gamma$-ray emitting blazars reside according to \cite{WGS_1}. Thus this is the most probable counterpart of the {\it Fermi}-detected $\gamma$-ray source. We compared the SED of Complex A to the average SED of {\it Fermi}-detected blazars \citep{Fermi_blazarseq}. The observed $\gamma$-ray flux density is much higher than those expected from the low-energy part of the SED. This is possibly due to variability since the radio, infrared, optical and $\gamma$-ray data were not observed at the same time.

The nature of Complexes B and C is still ambiguous. They are unlikely to be related to the blazar in Complex A. In that case one would expect to detect similar radio-emitting feature(s) on the other side of the blazar. If they are physically unrelated to Complex A, they can still belong to a single object, (i) being the two lobes of a radio galaxy or (ii) being the gravitationally lensed images of one background source. In the first case, the faint optical source and coincident infrared source located between the two radio-emitting features might be the host galaxy. However, the surrounding interstellar matter has to be quite asymmetric to cause an advancing lobe (Complex B) seen much closer to the host galaxy in projection than the receding one. In the second case, a cluster of galaxies are needed to act as a lens to provide the large image separation observed. However, there is no sign of such in the SDSS images. 

Intermediate-resolution deep radio observations may reveal additional faint radio-emitting features providing support for the scenario which explains Complexes B and C as lobes of a radio galaxy. Additional sensitive optical observation can shed light on the nature of the optical and infrared source located between the two radio Complexes B and C.

Since the radio source 4C$+$29.48 when imaged with sufficiently high resolution consists of three potentially unrelated radio sources seen in projection, we suggest the {\it Fermi} source 3FGL\,J1323.0$+$2942 to be associated with the blazar residing in Complex A and we suggest to name it J1323$+$2941A in subsequent additions of the {\it Fermi} catalog. Additionally, we propose to designate Complexes B and C as J1323$+$2941B and J1323$+$2941C, respectively. This would eliminate potential misinterpretations arising because the coordinates of 4C$+$29.48 are close to those of the middle component (Complex B), while we have shown that the $\gamma$-ray source is almost certainly the blazar J1323$+$2941A (Complex A) at $z=1.142$.

\bibliographystyle{aa} 
\bibliography{ref_gamma} 

\begin{acknowledgements}
This project was supported by the J\'anos Bolyai Research Scholarship of the Hungarian Academy of Sciences, by the Hungarian National Research Development and Innovation Office (OTKA NN110333), and by the China--Hungary Collaboration and Exchange Programme by the International Cooperation Bureau of the Chinese Academy of Sciences (CAS). T. A. acknowledges the grant of the Youth Innovation Promotion Association of CAS.

K\'EG wishes to thank G. Orosz for his help with the SED plotting.

We made use of the Astrogeo VLBI FITS image database (\url{http://astrogeo.org/vlbi_images}).

The National Radio Astronomy Observatory is a facility of the National Science Foundation operated under cooperative agreement by Associated Universities, Inc.

The European VLBI Network is a joint facility of independent European, African, Asian, and North American radio astronomy institutes. Scientific results from data presented in this publication are derived from the following EVN project code: EL034A.

This research has made use of the NASA/IPAC Extragalactic Database (NED) which is operated by the Jet Propulsion Laboratory, California Institute of Technology, under contract with the National Aeronautics and Space Administration.

Funding for SDSS-III has been provided by the Alfred P. Sloan Foundation, the Participating Institutions, the National Science Foundation, and the U.S. Department of Energy Office of Science. The SDSS-III web site is http://www.sdss3.org/. SDSS-III is managed by the Astrophysical Research Consortium for the Participating Institutions of the SDSS-III Collaboration including the University of Arizona, the Brazilian Participation Group, Brookhaven National Laboratory, Carnegie Mellon University, University of Florida, the French Participation Group, the German Participation Group, Harvard University, the Instituto de Astrofisica de Canarias, the Michigan State/Notre Dame/JINA Participation Group, Johns Hopkins University, Lawrence Berkeley National Laboratory, Max Planck Institute for Astrophysics, Max Planck Institute for Extraterrestrial Physics, New Mexico State University, New York University, Ohio State University, Pennsylvania State University, University of Portsmouth, Princeton University, the Spanish Participation Group, University of Tokyo, University of Utah, Vanderbilt University, University of Virginia, University of Washington, and Yale University. 

This publication makes use of data products from the Wide-field Infrared Survey Explorer, which is a joint project of the University of California, Los Angeles, and the Jet Propulsion Laboratory/California Institute of Technology, funded by the National Aeronautics and Space Administration.

This work has made use of data from the European Space Agency (ESA) mission {\it Gaia} (\url{https://www.cosmos.esa.int/gaia}), processed by the {\it Gaia} Data Processing and Analysis Consortium (DPAC, \url{https://www.cosmos.esa.int/web/gaia/dpac/consortium}). Funding for the DPAC has been provided by national institutions, in particular the institutions participating in the {\it Gaia} Multilateral Agreement.

Part of this work is based on archival data, and online services provided by the Space Science Data Center - ASI.

\end{acknowledgements}

\end{document}